\documentclass[fleqn, prb,twocolumn,preprintnumbers,amsmath,amssymb]{revtex4}

\usepackage{graphicx}
\usepackage{dcolumn}
\usepackage{bm}

\begin{document}


\title{Exciton and hole spin dynamics in ZnO}

\author{D. Lagarde, A. Balocchi, P. Renucci, H. Carr\`ere,  F. Zhao, T. Amand, X. Marie}
 \email{marie@insa-toulouse.fr}
\affiliation{Universit\'e de Toulouse; INSA, UPS; LPCNO, 135 avenue de Rangueil, F-31077 Toulouse, France and CNRS; LPCNO, F-31077 Toulouse, France
}

\author{Z.X. Mei, X.L. Du, Q.K. Xue}
\affiliation{Institute of Physics, Academy of Sciences and National Center for Nano-Science and Technology, Beijing 100080, China}%

\date{\today}

\begin{abstract}
The carrier spin dynamics in ZnO is investigated by time-resolved optical orientation experiments. We evidence a clear circular polarization of the donor-bound exciton luminescence in both ZnO epilayer and non-intentionally doped bulk ZnO. This allows us to measure the localized hole spin relaxation time. We find $\tau^{s}_h$$\sim$350 ps at T=1.7 K in the ZnO epilayer. The strong energy and temperature dependences of the photoluminescence polarization dynamics are well explained by the fast free exciton spin relaxation time and the ionization of bound excitons. 
\end{abstract}
%
\maketitle

Wide bandgap oxide semiconductor ZnO and its related heterostructures have raised substantial interest in the optoelectronics-oriented research field in the blue/ultra-violet range \cite{Klingshirn}. Besides, with a small spin-orbit coupling and a very large exciton binding energy, ZnO represents a potential candidate for room-temperature spintronic applications. Room-temperature ferromagnetism has also been predicted for dilute-Mn-doped ZnO \cite{Dietl}. However, only few measurements on the carrier spin dynamics in bulk or even nanostructured ZnO have been published to date. Ghosh \textit{et al.} have investigated the electron spin properties in \textit{n}-type ZnO structures \cite{Ghosh} and found an electron spin relaxation time varying from 20 ns to 190 ps when the temperature increases from T=10 to 280 K. Room-temperature electron spin relaxation as long as 25 ns has also been measured by electron paramagnetic resonance spectroscopy in colloidal \textit{n}-doped ZnO quantum dots \cite{Liu}. 

To the best of our knowledge, neither the exciton nor the hole spin dynamics in ZnO have been measured yet. Indeed, two experimental issues arise: (i) the small value of the spin-orbit coupling energy (9-16 meV\cite{Reynolds_99, Lambrecht}) imposes strictly- or quasi-resonant optical excitation conditions \cite{Chen} in the near-UV to create an exciton spin polarization; (ii) the direct measurement of the free hole spin relaxation by pump-probe measurements would require the fabrication of stable \textit{p}-doped samples \cite{Ohno, Syperek}, which remains a challenge in ZnO \cite{Ozgur}. In order to investigate the hole spin dynamics in ZnO, we have studied the polarization properties of the exciton bound to neutral donors. Since this complex consists of a singlet of electrons and a hole, its spin-polarization is directly determined by the orientation of the hole bound in the complex \cite{Planel, Tribollet, Laurent}. 

ZnO crystallizes in the wurtzite phase where the hexagonal crystal field $\Delta_{cr}$ and the spin orbit coupling $\Delta_{so}$ give rise to three doubly-degenerated valence bands, labelled A, B, C. The optical selection rules and oscillator strengths impose that only the transitions from the A and B valence bands are optically allowed when the light propagates along the \textit{c}-axis of the crystal\cite{Reynolds_99, Planel, Lambrecht}. 

We present in this paper a detailed investigation of the optical orientation of excitons and holes in ZnO bulk and epilayer samples. By time-resolved photoluminescence (PL) experiments, we evidence (i) the fast free exciton spin relaxation time ($\tau^{s}_{_{FX}}$$<$10 ps) and (ii) the hole spin relaxation time (up to $\tau^{s}_{h}\sim$350 ps) in the donor-bound exciton complex.  

The samples under investigation have been grown by two different methods. This gives us the opportunity to generalize our results, independently from the growth conditions. Sample I is a high quality, nominally undoped, 1.1 $\mu$m thick epilayer grown on a $\alpha$-sapphire (0001) substrate using an rf-plasma-assisted MBE system. The nitridation of the substrate surface has provided the growth of the single-domain Zn-polar ZnO epitaxial sample \cite{Du}. Sample II consists of a commercial bulk ZnO substrate submitted to an O-plasma treatment to improve the crystal quality of the surface. 

Reflectivity experiments have been performed using a tungsten-halogen lamp focused onto the sample at normal incidence. In the time-resolved PL experiments, the excitation source is a mode-locked frequency-doubled Ti:Sa laser, with a 1.5 ps pulse width, a wavelength in the 350-380 nm range and a repetition frequency of 80 MHz. The laser beam propagating along the growth \textit{c}-axis is focused onto the sample to a 100 $\mu$m diameter spot with an average power $P_{exc}$=0.5 mW\cite{Puissance}. The PL intensity is dispersed by an imaging spectrometer with a spectral resolution of 0.5 meV. The temporal and spectral properties of the signal are recorded by a S20 photocathode streak camera with an overall time-resolution of 8 ps. The excitation laser is circularly-polarized ($\sigma^+$) and the resulting PL circular polarization $P_{c}$ is calculated as: $P_{c}$=$(I^+ - I^-)/(I^+ + I^-)$. Here, $I^+$ and $I^-$ are the PL intensity components co- and counter-polarized to the ($\sigma^+$) excitation laser.

\begin{figure}
\includegraphics[width=0.5\textwidth]{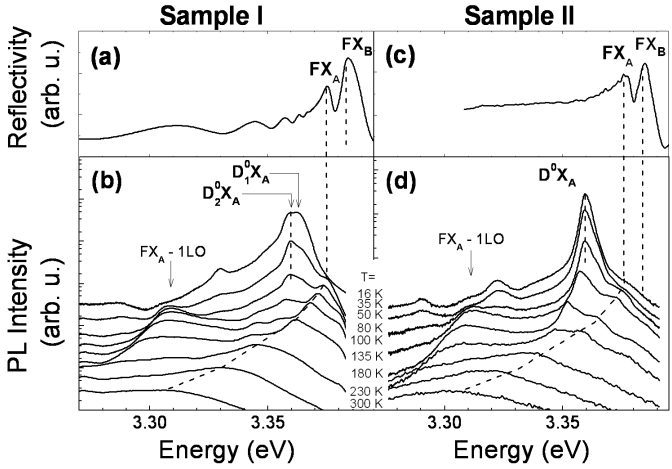}
\caption{(a)-(c) Reflectivity at T=16 K and (b)-(d) time-integrated photoluminescence spectra from T=16 to 300 K of sample I and II, respectively. Dotted lines are guides to the eye.}
\label{fig1}
\end{figure}
Figure~\ref{fig1} (a) and (c) present the reflectivity spectra at T=16 K of sample I and II, respectively. We clearly observe the absorption features corresponding to the free excitons $FX_A$ and $FX_B$ $ $ \cite{Reynolds_99}. Figure~\ref{fig1} (b) and (d) display the dependence of the time-integrated PL spectra as a function of temperature under nonresonant conditions ($E_{exc}$=3.450 eV). The ZnO luminescence is characterized by a rich structure of excitonic lines \cite{Reynolds_98, Meyer} and is dominated in sample I by two main lines labelled $D_{1}^{0}X_A$ and $D_{2}^{0}X_A$ at 3.363 meV and 3.359 meV, respectively. In sample II, the main line, labelled $D^{0}X_A$, is identified at 3.361 meV. The temperature evolution of the PL intensity on these lines allows us to identify them as excitons bound to neutral donors, as observed previously by different groups \cite{Reynolds_98, Ozgur, Meyer, Magneto}. Additionally, the dependence of the free exciton energy $FX_A$ as a function of temperature is clearly resolved. This latter attribution is indeed confirmed by the reflectivity spectra. LO-phonon replica are also identified in the low energy part of the spectra with $E_{_{LO}}$=72 meV \cite{Meyer}. 

\begin{figure}
\includegraphics[width=0.5\textwidth]{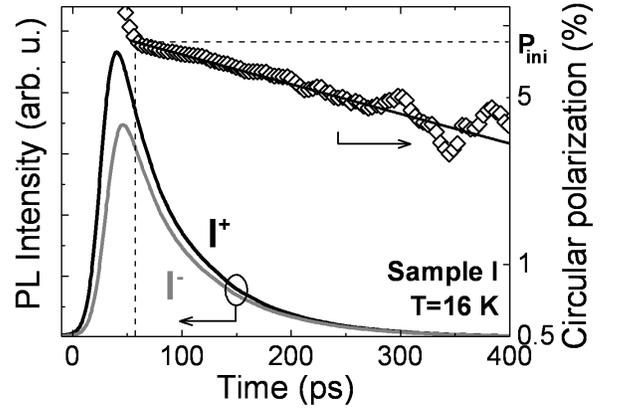}
\caption{Time evolution of the circular luminescence components ($I^+$) and ($I^-$) after a ($\sigma^+$)-polarized excitation at T=16 K in sample I. The circular polarization (right axis) is also plotted. The detection energy is set to the $D_{2}^{0}X_A$ energy position while the excitation laser is tuned to 11 meV above the detection.}
\label{fig2}
\end{figure}

Figure~\ref{fig2} presents the measured PL circular polarization dynamics following a circularly-polarized ($\sigma^+$) excitation of sample I when the detection is fixed on the $D_{2}^{0}X_A$ energy position, at T=16 K. The laser excitation energy is tuned to 11 meV above the detection but below the $FX_A$ energy. The initial polarization $P_{ini}$ is measured 60 ps after the laser excitation so that it is not impacted by the polarized laser light backscattered by the sample surface. As the total electron spin of the donor-bound exciton is zero, the measured PL circular polarization corresponds directly to the hole spin polarization. Assuming an exponential decay of the PL circular polarization dynamics $P_{c}$$\sim$exp(-t/$\tau^{s}_h$), we directly determine the hole spin relaxation time $\tau^{s}_h$$\sim$350 ps at T=16 K in sample I. 

\begin{figure}
\includegraphics[width=0.5\textwidth]{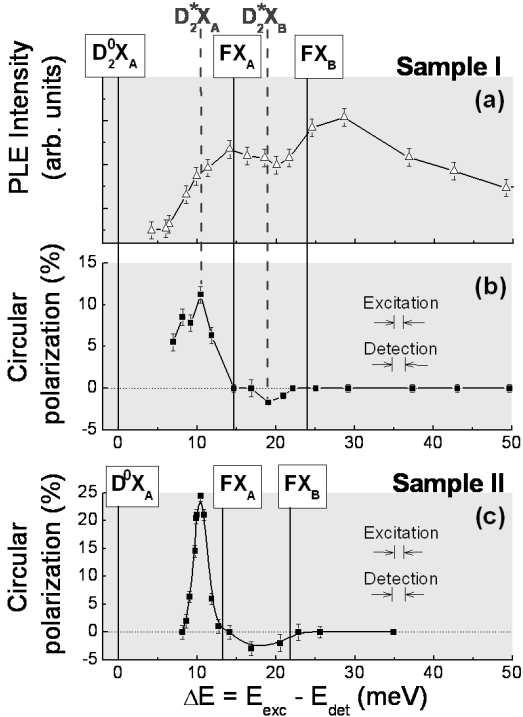}
\caption{ (a) Photoluminescence excitation (sample I) and (b) dependence of the circular polarization $P_{ini}$ detected on $D_{2}^{0}X_A$ (sample I) and (c) on $D^{0}X_A$  (sample II), as a function of the energy difference $\Delta$E between excitation and detection energies at T=16 K. Energy resolution for excitation and detection is indicated.}
\label{fig3}
\end{figure}

We have analysed the circular polarization $P_{ini}$ of the luminescence from the $D_{2}^{0}X_A$ line following a circularly-polarized ($\sigma^+$) excitation as a function of the laser excitation energy. Figure~\ref{fig3}(b) displays the spectral dependence of the $D_{2}^{0}X_A$ circular polarization $P_{ini}$ for sample I when the energy difference $\Delta$E between the excitation energy $E_{exc}$ and detection energy $E_{det}$ is varied from 8 meV to 50 meV. The corresponding time-integrated\cite{Integration} photoluminescence excitation (PLE) spectrum is also plotted (Fig.~\ref{fig3}(a)) with a clear signature of the $FX_A$ and $FX_B$ free excitons. Under quasi-resonant conditions, i.e. when the excitation energy is set below the free exciton $FX_A$ energy, we measure a significant PL circular polarization. The peak value $P_{ini}$=10 $\%$ is measured when the energy difference is $\Delta$E=11 meV. No polarization within our temporal resolution is measured when the excitation is resonant with the $FX_A$ line. The degree of polarization becomes then negative when the excitation energy is tuned between the $FX_A$ and $FX_B$ energies. This negative value, i.e. opposite to the helicity of the excitation laser, has been carefully checked and will be discussed later. When the excitation energy is higher than the free exciton $FX_B$ energy, no more circular polarization is measured, as expected by the optical selection rules. Similar results were obtained on sample II, as shown in Fig.~\ref{fig3}(c), with a peak value ($P_{ini}$$\sim$25 $\%$) when $E_{exc}$ is tuned 10 meV above the detection energy. 

\begin{figure}
\includegraphics[width=0.5\textwidth]{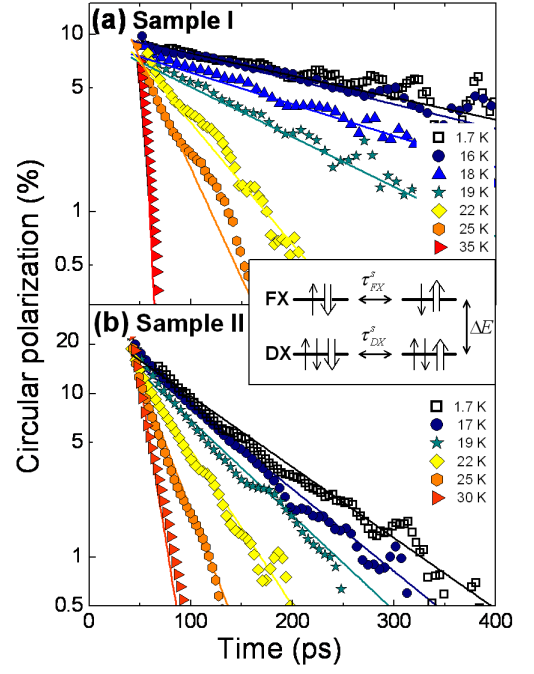}
\caption{Temperature dependence of the circular polarization detected (a) on $D_{2}^{0}X_A$ (sample I) and (b) on $D^{0}X_A$ (sample II). The excitation energy is $E_{exc}$=3.370 eV and $E_{exc}$=3.371 eV in sample I and II, respectively. The lines correspond to the results of the calculations (see text). Inset: schematic representation of the spin states for free and bound excitons,  $\uparrow,\downarrow$: electron spin; $\Uparrow,\Downarrow$:hole spin.}
\label{fig4}
\end{figure}

We now describe the time-resolved results obtained when exciting at the energy that resulted in a maximum polarization for each sample. The detection is fixed at the bound exciton emission lines i.e.~$D_{2}^{0}X_A$ (sample I) and $D^{0}X_A$ (sample II). Figure~\ref{fig4} shows the time evolution of the corresponding circular polarization as a function of temperature in the range 1.7 - 35 K. For T=1.7 K, the polarization decay time is $\tau^{s}_h$$\sim$350 ps and $\sim$100 ps on sample I and II, respectively. When the temperature increases by a few tens of kelvin, it drops drastically, i.e. $\tau^{s}_h$ is shorter than 8 ps when the temperature exceeds 30 K in both samples. 
This fast drop of the polarization decay time can not be attributed directly to the temperature dependence of the localized hole spin relaxation time, as shown below. First, we observe that its activation energy (not shown) is close to the binding energy of the donor-bound excitons i.e. $\sim$15 meV for $D_{2}^{0}X_A$ of sample I and $\sim$13 meV for $D^{0}X_A$ of sample II. Secondly, we did not manage to observe any circular polarization of the PL within our temporal resolution under quasi-resonant excitation of the free exciton $FX_A$. From this measurement, we can infer that the free exciton spin relaxation time $\tau^{s}_{_{FX}}$$<$10 ps, which is indeed consistent with the one measured in wurtzite GaN ($\tau^{s}_{_{FX}}$$\sim$1 ps)\cite{Ishiguro, Gilliot}, a material with similar electronic structure. Thus, we interpret the strong temperature dependence of the bound exciton circular polarization measured in both samples as a consequence of the quasi-equilibrium population distribution between the free and bound states. When the temperature increases, one electron and the hole of the donor-bound exciton can gain sufficient energy to be released as a free exciton. In the free phase, the optically active (J=1) exciton can efficiently loose its spin orientation and, when it is trapped again on the donor, both the electron and the hole have lost their initial spin orientation. This mechanism, alone, could explain the temperature dependence observed in Fig.~\ref{fig4}. 

In order to have a better description of the mechanism, we use simple rate equations to characterize the evolution as a function of time and temperature of the free and bound exciton populations and their spin polarization (see insets in Fig.~4 and Fig.~5). Two main simplifying assumptions allow us to resolve analytically the system. 

(i) The injected carrier density is low compared to the density of donors. This approximation is justified by the linear increase of the $D^{0}X_A$ PL intensity as a function of excitation power $P_{exc}$ in the investigated range under non-resonant excitation conditions. We find\cite{Relaxation}:
\begin{footnotesize}
\begin{eqnarray}
 \left\{ \begin{array}{ll}
 \dot{n}_{_{J=1}} = - \Big( \frac{1}{\tau_{_{FX}}}+\frac{1}{2\tau_{capt}}\Big)n_{_{J=1}} +\frac{1}{\tau_{ion}}n_{_{DX}} \\
 \dot{n}_{_{J=2}} = - \frac{1}{2\tau_{capt}}n_{_{J=2}} +\frac{1}{\tau_{ion}}n_{_{DX}} \\
 \dot{n}_{_{DX}} = - \Big( \frac{1}{\tau_{_{DX}}}+\frac{2}{\tau_{ion}}\Big)n_{_{DX}} +\frac{1}{2\tau_{capt}}(n_{_{J=1}}+n_{_{J=2}})
\end{array} \right.
\end{eqnarray}
\begin{eqnarray}
 \left\{ \begin{array}{ll}
  \dot{S}_{_{J=1}} = - \Big( \frac{1}{\tau^{s}_{_{FX}}}+\frac{1}{\tau_{_{FX}}}+\frac{1}{2\tau_{capt}}\Big)S_{_{J=1}} +\frac{1}{\tau_{ion}}S_{_{DX}} \\
  \dot{S}_{_{J=2}} = - \frac{1}{2\tau_{capt}}S_{_{J=2}} +\frac{1}{\tau_{ion}}S_{_{DX}} \\
  \dot{S}_{_{DX}} = - \Big( \frac{1}{\tau^{s}_{h}}+\frac{1}{\tau_{_{DX}}}+\frac{2}{\tau_{ion}}\Big)S_{_{DX}} +\frac{1}{2\tau_{capt}}(S_{_{J=1}}+S_{_{J=2}}) \\
\end{array} \right.
\end{eqnarray} 
\end{footnotesize}
where $n_{_{J=1}}$ ($n_{_{J=2}}$) and $S_{_{J=1}}$ ($S_{_{J=2}}$) are the population and pseudo-spin densities of optically active J=1 (dark J=2) free excitons, respectively; $n_{_{DX}}$ and $S_{_{DX}}$ are the population and spin densities of bound excitons, respectively; $\tau_{_{FX}}$ ($\tau_{_{DX}}$) are the intrinsic recombination times of the free (bound) exciton; $\tau^{s}_{_{FX}}$ and $\tau^{s}_{h}$ are the spin relaxation time of the optically active free exciton and the localized hole, respectively. The capture and ionization times of excitons are given by $\tau_{capt}$ and $\tau_{ion}$, respectively (see inset of Fig.~5)\cite{Ionization}. As the energy separation of the free and bound excitons is $\Delta E\sim$15 meV in sample I and 13 meV in sample II, the ionization of bound excitons can be thermally activated via acoustic phonons. We can assume that $\tau_{capt}$ and $\tau_{ion}$ are linked by the relation: 
\begin{equation}
\tau_{ion}(T)=\tau_{capt}\frac{N_{_{D}}}{N_{_{FX}}(T)}\mathrm{exp}(\frac{\Delta E}{k_{B}T})
\end{equation}
where $N_{_D}$ and $N_{_{FX}}$ are the donor concentration and the equivalent density of free exciton states, respectively. In a simple approach, $N_{_{FX}}$ can be written as: $\mathrm{N_{_{FX}}(T)}$=$\left( \frac{\mathrm{2\pi M^{*}kT}}{h^2}\right)^{3/2}$ where $M^*$=0.85$m_0$ is the effective exciton mass \cite{Lambrecht}. 

(ii) The free exciton capture is shorter than the ionization of bound excitons and than the FX recombination ($\tau_{capt}$$<$$\tau_{ion}$, $\tau_{_{FX}}$). It comes that, after a short transient corresponding to the fast capture of the free excitons, the FX population and spin densities are directly governed by the ones on the donor levels, so that we can write:
\begin{footnotesize}
\begin{equation}
n_{_{J=1}}\sim\frac{1}{\tau_{ion}}\frac{2\tau_{_{FX}}\tau_{capt}}{\tau_{_{FX}}+2\tau_{capt}}n_{_{DX}}
\end{equation}
\begin{equation}
n_{_{J=2}}\sim\frac{2\tau_{capt}}{\tau_{ion}}n_{_{DX}}
\end{equation}
\begin{equation}
S_{_{J=1}}\sim\frac{1}{\tau_{ion}}\frac{2\tau^{s}_{_{FX}}\tau_{capt}}{\tau^{s}_{_{FX}}+2\tau_{capt}}S_{_{DX}}
\end{equation}
\begin{equation}
S_{_{J=2}}\sim\frac{2\tau_{capt}}{\tau_{ion}}S_{_{DX}}
\end{equation}
\end{footnotesize}
Fig.~5 illustrates this mechanism, comparing the time evolution of the free (J=1) and bound exciton luminescence on sample I at T=20 K. The $FX_A$ PL decay is fitted by a bi-exponential law, where the first time gives a good approximation of the capture time ($\tau_{capt}$$\sim$10 ps) and the second time directly corresponds to the monoexponential PL decay of the $D^{0}X_A$ line ($\tau_{_{DX}}$$\sim$60 ps). Similar results were found on sample I. 

Thus, we obtain first order differential equations to describe the bound exciton population and spin polarization, with the characteristic times $\tau^{PL}_{_{DX}}$ and $\tau^{s}_{_{DX}}$, respectively given by:
\begin{footnotesize}
\begin{equation}
  \frac{1}{\tau^{PL}_{_{DX}}(T)}\sim \frac{1}{\tau_{_{DX}}}+ \frac{1}{\tau_{ion}(T)}\Big(1-\frac{\tau_{_{FX}}}{\tau_{_{FX}}+2\tau_{capt}}\Big)
\end{equation}
\begin{equation}
  \frac{1}{\tau^{s}_{_{DX}}(T)}\sim \frac{1}{\tau^{s}_{h}}+ \frac{1}{\tau_{ion}(T)} \Big(\frac{\tau_{_{FX}}}{\tau_{_{FX}}+2\tau_{capt}}-\frac{\tau^{s}_{_{FX}}}{\tau^{s}_{_{FX}}+2\tau_{capt}}\Big)
\end{equation}
\end{footnotesize}
\begin{figure}
\includegraphics[width=0.5\textwidth]{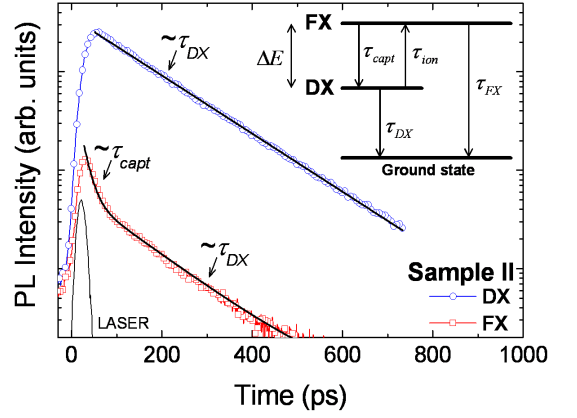}
\caption{Time evolution of the total PL intensity detected on the $FX_A$ and $D^{0}X_A$ lines of sample II at T=20 K. The full lines are a fit of the PL decays. The excitation energy is $E_{exc}$=3.450 eV. Inset: Schematic representation of the excitonic recombination and thermalization paths (see text).}
\label{fig5}
\end{figure}

The fitting procedure is the following. First, we use nonresonant PL experiments ($E_{exc}$$>$$FX_A$) at T=1.7 K to measure the recombination times of the bound excitons (not shown). We find $\tau_{_{DX}}$$\sim$60 ps for sample I and $\tau_{_{DX}}$$\sim$160 ps for sample II. Using Eq.(8), we then calculate the temporal evolution of the total luminescence intensity of the bound excitons as a function of temperature. The fitting parameters are $\tau{_{FX}}$ and $N_{_D}$; a very good agreement between the experiments and the calculations is achieved for $\tau_{_{FX}}$$\sim$50 ps and $N_{_D}$$\sim$5$\cdot$$10^{15}$ $cm^{-3}$. This latter is a reasonable value compared to the density of residual \textit{n}-doping in sample I. 

Finally, we use the same set of parameters in the calculations of the temperature-dependent circular polarization dynamics of the bound excitons using Eq.(9). The fitting parameters are the spin relaxation time of the free exciton ($\tau^{s}_{_{FX}}$$\sim$1 ps in both samples) and the hole spin lifetime $\tau^{s}_{h}$. Note that we reproduce the experimental data with a very good agreement in both samples (\textit{cf} Fig.~\ref{fig4}) considering $\tau^{s}_{h}$ as independent of the temperature in the range 1.7 - 35 K. The hole localization probably makes its spin orientation only weakly sensitive to the temperature in this range. We find $\tau^{s}_{h}\sim$350 ps and $\sim$100 ps for sample I and II, respectively. The discrepancy between the two measured values in the epilayer and bulk samples could be attributed to the presence of compressive strain in the former \cite{Ghosh, Baylac}. Besides, let us note that experiments at T=1.7 K confirm those results since the measured $\tau^{s}_{_{DX}}$ directly corresponds to $\tau^{s}_{h}$ as the ionization at this temperature is negligible. 

To complete the analysis, we now discuss the results displayed on Fig.~\ref{fig3}(b) and (c). On both samples, the maximum polarization value is detected on the bound exciton for a selective excitation energy $\sim$11 meV above the neutral-donor energy. This corresponds to the resonant excitation of an excited state (labelled $D^{*}_{2}X_A$ in Fig.~3 (b)) of the donor-bound exciton\cite{Gutowski, Morhain}. The absence of a measurable polarization when the excitation energy is resonant with the free excitons, is consistent with the fast free exciton spin relaxation time ($\tau^{s}_{_{FX}}$$<$10 ps): the hole spin orientation is lost before the capture of the exciton by the donor. Regarding the negative polarization, measured when the excitation energy is tuned between the $FX_A$ and $FX_B$ excitons, a possible explanation is that this level corresponds to an excited state (labelled $D^{*}_{2}X_B$ in Fig.~3) of a donor-bound exciton constructed with a hole from the B valence band ($D_{0}X_B$). The relaxation to the $D^{0}X_A$ level can be achieved by the emission of an acoustic phonon: the momentum difference is compensated by the orbital part of the exciton wavefunction, leaving unchanged the spin contribution. Thus, the resulting PL polarization is negative (i.e. counter-polarized to the laser excitation) \cite{Sham, Barrau}. 

In conclusion, we have measured hole spin relaxation times as long as $\tau^{s}_h$$\sim$350 ps at T=1.7 K when the holes are localized by a donor potential in ZnO. The spectral dependence of the measured PL polarization demonstrates the fast spin relaxation of the free excitons and the implication of holes from both the A and B valence bands. These experimental data should stimulate theoretical investigations of the spin relaxation processes in ZnO.

The authors are grateful to the EADS Research foundation, Institut Universitaire de France and AFCRST for financial support.


\begin{thebibliography}{32}
\expandafter\ifx\csname natexlab\endcsname\relax\def\natexlab#1{#1}\fi
\expandafter\ifx\csname bibnamefont\endcsname\relax
  \def\bibnamefont#1{#1}\fi
\expandafter\ifx\csname bibfnamefont\endcsname\relax
  \def\bibfnamefont#1{#1}\fi
\expandafter\ifx\csname citenamefont\endcsname\relax
  \def\citenamefont#1{#1}\fi
\expandafter\ifx\csname url\endcsname\relax
  \def\url#1{\texttt{#1}}\fi
\expandafter\ifx\csname urlprefix\endcsname\relax\def\urlprefix{URL }\fi
\providecommand{\bibinfo}[2]{#2}
\providecommand{\eprint}[2][]{\url{#2}}

\bibitem[{\citenamefont{Klingshirn et~al.}(2006)\citenamefont{Klingshirn,
  Priller, Decker, Br\"uckner, Kalt, Hauschild, Zeller, Waag, Bakin, Wehmann, 
  Thonke, Sauer, Kling, Reuss and Kirchner}}]{Klingshirn}
\bibinfo{author}{\bibfnamefont{C.}~\bibnamefont{Klingshirn}},
  \bibinfo{author}{\bibfnamefont{H.}~\bibnamefont{Priller}},
  \bibinfo{author}{\bibfnamefont{M.}~\bibnamefont{Decker}},
  \bibinfo{author}{\bibfnamefont{J.}~\bibnamefont{Br\"uckner}},
  \bibinfo{author}{\bibfnamefont{H.}~\bibnamefont{Kalt}},
  \bibinfo{author}{\bibfnamefont{R.}~\bibnamefont{Hauschild}},
  \bibinfo{author}{\bibfnamefont{J.}~\bibnamefont{Zeller}},
  \bibinfo{author}{\bibfnamefont{A.}~\bibnamefont{Waag}},
  \bibinfo{author}{\bibfnamefont{A.} \bibnamefont{Bakin}},
  \bibinfo{author}{\bibfnamefont{H.}~\bibnamefont{Wehmann}},
  \textit{\bibnamefont{et~al.}}, \emph{\bibinfo{title}{Excitonic properties of ZnO}}, 
  \bibinfo{publisher}{Springer Berlin-Heidelberg}, \bibinfo{volume}{45}, \bibinfo{year}{2006}).
  
\bibitem[{\citenamefont{Dietl et~al.}(2000)\citenamefont{Dietl,
  Ohno, Matsukura, Cibert, and Ferrand}}]{Dietl}
\bibinfo{author}{\bibfnamefont{T.}~\bibnamefont{Dietl}},
  \bibinfo{author}{\bibfnamefont{H.}~\bibnamefont{Ohno}},
  \bibinfo{author}{\bibfnamefont{F.}~\bibnamefont{Matsukura}},
  \bibinfo{author}{\bibfnamefont{J.} \bibnamefont{Cibert}}, \bibnamefont{and}
  \bibinfo{author}{\bibfnamefont{D.} \bibnamefont{Ferrand}},
  \bibinfo{journal}{Science} \textbf{\bibinfo{volume}{287}},
  \bibinfo{pages}{1019} (\bibinfo{year}{2000}). 
  
\bibitem[{\citenamefont{Ghosh et~al.}(2005)\citenamefont{Ghosh, Sih,
  Lau, Awschalom, Bae, Wang, Vaidya, and Chapline}}]{Ghosh}
\bibinfo{author}{\bibfnamefont{S.}~\bibnamefont{Ghosh}},
  \bibinfo{author}{\bibfnamefont{V.}~\bibnamefont{Sih}},
  \bibinfo{author}{\bibfnamefont{W.~H.}~\bibnamefont{Lau}},
  \bibinfo{author}{\bibfnamefont{D.~D.}~\bibnamefont{Awschalom}},
  \bibinfo{author}{\bibfnamefont{S.-Y.}~\bibnamefont{Bae}},
  \bibinfo{author}{\bibfnamefont{S.}~\bibnamefont{Wang}},
  \bibinfo{author}{\bibfnamefont{S.}~\bibnamefont{Vaidya}}, \bibnamefont{and}
  \bibinfo{author}{\bibfnamefont{G.}~\bibnamefont{Chapline}}, 
  \bibinfo{journal}{Appl. Phys. Lett.} \textbf{\bibinfo{volume}{86}}, \bibinfo{pages}{232507}
  (\bibinfo{year}{2005}).  

\bibitem[{\citenamefont{Liu et~al.}(2007)\citenamefont{Liu, Whitaker,
  Smith, Kittilstved, Robinson, and Gamelin}}]{Liu}
\bibinfo{author}{\bibfnamefont{W.~K.}~\bibnamefont{Liu}},
  \bibinfo{author}{\bibfnamefont{K.~M.}~\bibnamefont{Whitaker}},
  \bibinfo{author}{\bibfnamefont{A.~L.}~\bibnamefont{Smith}},
  \bibinfo{author}{\bibfnamefont{K.~R.}~\bibnamefont{Kittilstved}},
  \bibinfo{author}{\bibfnamefont{B.~H.}~\bibnamefont{Robinson}}, \bibnamefont{and}
  \bibinfo{author}{\bibfnamefont{D.~R.}~\bibnamefont{Gamelin}},
  \bibinfo{journal}{Phys. Rev. Lett.} \textbf{\bibinfo{volume}{98}},
  \bibinfo{pages}{186804} (\bibinfo{year}{2007}).  
  
\bibitem[{\citenamefont{Reynolds et~al.}(1999)\citenamefont{Reynolds,
  Look, Jogai, Litton, Cantwell, and Harsch}}]{Reynolds_99}
\bibinfo{author}{\bibfnamefont{D.~C.}~\bibnamefont{Reynolds}},
  \bibinfo{author}{\bibfnamefont{D.~C.}~\bibnamefont{Look}},
  \bibinfo{author}{\bibfnamefont{B.} \bibnamefont{Jogai}},
  \bibinfo{author}{\bibfnamefont{C.~W.}~\bibnamefont{Litton}},
  \bibinfo{author}{\bibfnamefont{G.} \bibnamefont{Cantwell}}, \bibnamefont{and}
  \bibinfo{author}{\bibfnamefont{W.~C.} \bibnamefont{Harsch}},
  \bibinfo{journal}{Phys. Rev. B} \textbf{\bibinfo{volume}{60}},
  \bibinfo{pages}{2340} (\bibinfo{year}{1999}). 
  
\bibitem[{\citenamefont{Lambrecht et~al.}(2002)\citenamefont{Lambrecht, Rodina,
  Limpijumnong, Segall, and Meyer}}]{Lambrecht}
\bibinfo{author}{\bibfnamefont{W.~R.L.}~\bibnamefont{Lambrecht}},
  \bibinfo{author}{\bibfnamefont{A.~V.}~\bibnamefont{Rodina}},
  \bibinfo{author}{\bibfnamefont{S.}~\bibnamefont{Limpijumnong}},
  \bibinfo{author}{\bibfnamefont{B.}~\bibnamefont{Segall}}, \bibnamefont{and}
  \bibinfo{author}{\bibfnamefont{B.~K.}~\bibnamefont{Meyer}},
  \bibinfo{journal}{Phys. Rev. B} \textbf{\bibinfo{volume}{65}},
  \bibinfo{pages}{075207} (\bibinfo{year}{2002}).  
  
\bibitem[{\citenamefont{Chen et~al.}(2008)\citenamefont{Chen, Buyanova,
  Murayama, Furuta, Oka, Norton, Pearton, Osinsky, and Dong}}]{Chen}
\bibinfo{author}{\bibfnamefont{W.~M.}~\bibnamefont{Chen}},
  \bibinfo{author}{\bibfnamefont{I.~A.}~\bibnamefont{Buyanova}},
  \bibinfo{author}{\bibfnamefont{A.}~\bibnamefont{Murayama}},
  \bibinfo{author}{\bibfnamefont{T.}~\bibnamefont{Furuta}},
  \bibinfo{author}{\bibfnamefont{Y.}~\bibnamefont{Oka}},
  \bibinfo{author}{\bibfnamefont{D.~P.}~\bibnamefont{Norton}},
  \bibinfo{author}{\bibfnamefont{S.~J.}~\bibnamefont{Pearton}},
  \bibinfo{author}{\bibfnamefont{A.}~\bibnamefont{Osinsky}}, \bibnamefont{and}
  \bibinfo{author}{\bibfnamefont{J.~W.}~\bibnamefont{Dong}},
  \bibinfo{journal}{Appl. Phys. Lett.} \textbf{\bibinfo{volume}{92}},
  \bibinfo{pages}{092103} (\bibinfo{year}{2008}) ;   
  \bibinfo{author}{\bibfnamefont{D.}~\bibnamefont{Lagarde}},
  \bibinfo{author}{\bibfnamefont{L.}~\bibnamefont{Lombez}},
    \bibinfo{author}{\bibfnamefont{A.}~\bibnamefont{Balocchi}},
    \bibinfo{author}{\bibfnamefont{P.}~\bibnamefont{Renucci}},
      \bibinfo{author}{\bibfnamefont{T.}~\bibnamefont{Amand}},
        \bibinfo{author}{\bibfnamefont{X.}~\bibnamefont{Marie}},
  \bibinfo{author}{\bibfnamefont{Z.~X.}~\bibnamefont{Mei}},
  \bibinfo{author}{\bibfnamefont{X.~L.}~\bibnamefont{Du}}, \bibnamefont{and}
  \bibinfo{author}{\bibfnamefont{Q.~K.}~\bibnamefont{Xue}},
  \bibinfo{journal}{Phys. Stat. Sol. c} \textbf{\bibinfo{volume}{4}},
  \bibinfo{pages}{472} (\bibinfo{year}{2007}).  
                
\bibitem[{\citenamefont{Hu et~al.}(2005)\citenamefont{Hu,
  Morita, Sanada, Matsuzaka, Ohno, and Ohno}}]{Ohno}
\bibinfo{author}{\bibfnamefont{C.~Y.} \bibnamefont{Hu}},
  \bibinfo{author}{\bibfnamefont{K.} \bibnamefont{Morita}},
    \bibinfo{author}{\bibfnamefont{H.} \bibnamefont{Sanada}},
    \bibinfo{author}{\bibfnamefont{S.} \bibnamefont{Matsuzaka}},
    \bibinfo{author}{\bibfnamefont{Y.} \bibnamefont{Ohno}},
  \bibnamefont{and} \bibinfo{author}{\bibfnamefont{H.}~\bibnamefont{Ohno}},
  \bibinfo{journal}{Phys. Rev. B} \textbf{\bibinfo{volume}{72}},
  \bibinfo{pages}{121203(R)} (\bibinfo{year}{2005}). 
     
\bibitem[{\citenamefont{Syperek et~al.}(2007)\citenamefont{Syperek,
  Yakovlev, Greilich, Misiewicz, Bayer, Reuter, and Wieck}}]{Syperek}
\bibinfo{author}{\bibfnamefont{M.} \bibnamefont{Syperek}},
  \bibinfo{author}{\bibfnamefont{D.~R.} \bibnamefont{Yakovlev}},
    \bibinfo{author}{\bibfnamefont{A.} \bibnamefont{Greilich}},
    \bibinfo{author}{\bibfnamefont{J.} \bibnamefont{Misiewicz}},
    \bibinfo{author}{\bibfnamefont{M.} \bibnamefont{Bayer}},
    \bibinfo{author}{\bibfnamefont{D.} \bibnamefont{Reuter}},
  \bibnamefont{and} \bibinfo{author}{\bibfnamefont{A.~D.}~\bibnamefont{Wieck}},
  \bibinfo{journal}{Phys. Rev. Lett.} \textbf{\bibinfo{volume}{99}},
  \bibinfo{pages}{187401} (\bibinfo{year}{2007}).  
  
\bibitem[{\citenamefont{\"Ozg\"ur et~al.}(1999)\citenamefont{\"Ozg\"ur,
  Alivov, Liu, Teke, Reshchikov, Do\u gan, Avrutin, Cho, and Morko\c c}}]{Ozgur}
\bibinfo{author}{\bibfnamefont{\"U}~\bibnamefont{\"Ozg\"ur}},
  \bibinfo{author}{\bibfnamefont{Ya.~I.}~\bibnamefont{Alivov}},
  \bibinfo{author}{\bibfnamefont{C.} \bibnamefont{Liu}},
  \bibinfo{author}{\bibfnamefont{A.}~\bibnamefont{Teke}},
  \bibinfo{author}{\bibfnamefont{M.~A.}~\bibnamefont{Reshchikov}},
  \bibinfo{author}{\bibfnamefont{S.}~\bibnamefont{Do\u gan}},
  \bibinfo{author}{\bibfnamefont{V.}~\bibnamefont{Avrutin}},
  \bibinfo{author}{\bibfnamefont{S.~J.} \bibnamefont{Cho}}, \bibnamefont{and}
  \bibinfo{author}{\bibfnamefont{H.} \bibnamefont{Morko\c c}},
  \bibinfo{journal}{Phys. Rev. B} \textbf{\bibinfo{volume}{60}},
  \bibinfo{pages}{2340} (\bibinfo{year}{1999}).    
  
\bibitem[{\citenamefont{Bonnot et~al.}(1974)\citenamefont{Bonnot,
  Planel, and Benoit \`a la Guillaume}}]{Planel}
\bibinfo{author}{\bibfnamefont{A.} \bibnamefont{Bonnot}},
  \bibinfo{author}{\bibfnamefont{R.} \bibnamefont{Planel}},
  \bibnamefont{and} \bibinfo{author}{\bibfnamefont{C.}~\bibnamefont{Benoit \`a la Guillaume}},
  \bibinfo{journal}{Phys. Rev. B} \textbf{\bibinfo{volume}{9}},
  \bibinfo{pages}{690} (\bibinfo{year}{1974}). 
  
\bibitem[{\citenamefont{Tribollet et~al.}(2007)\citenamefont{Tribollet,
  Aubry, Karczewski, Sermage, Bernardot, Testelin, and Chamarro}}]{Tribollet}
\bibinfo{author}{\bibfnamefont{J.}~\bibnamefont{Tribollet}},
  \bibinfo{author}{\bibfnamefont{E.} \bibnamefont{Aubry}},
    \bibinfo{author}{\bibfnamefont{G.} \bibnamefont{Karczewski}},
    \bibinfo{author}{\bibfnamefont{B.} \bibnamefont{Sermage}},
    \bibinfo{author}{\bibfnamefont{F.} \bibnamefont{Bernardot}},
    \bibinfo{author}{\bibfnamefont{C.} \bibnamefont{Testelin}},
  \bibnamefont{and} \bibinfo{author}{\bibfnamefont{M.}~\bibnamefont{Chamarro}},
  \bibinfo{journal}{Phys. Rev. B} \textbf{\bibinfo{volume}{75}},
  \bibinfo{pages}{205304} (\bibinfo{year}{2007}).  
  
\bibitem[{\citenamefont{Laurent et~al.}(2005)\citenamefont{Laurent,
  Eble, Krebs, Lema\^itre, Urbaszek, Marie, Amand, and Voisin}}]{Laurent}
\bibinfo{author}{\bibfnamefont{S.}~\bibnamefont{Laurent}},
  \bibinfo{author}{\bibfnamefont{B.} \bibnamefont{Eble}},
    \bibinfo{author}{\bibfnamefont{O.} \bibnamefont{Krebs}},
    \bibinfo{author}{\bibfnamefont{A.} \bibnamefont{Lema\^itre}},
    \bibinfo{author}{\bibfnamefont{B.} \bibnamefont{Urbaszek}},
    \bibinfo{author}{\bibfnamefont{X.} \bibnamefont{Marie}},
    \bibinfo{author}{\bibfnamefont{T.} \bibnamefont{Amand}},
  \bibnamefont{and} \bibinfo{author}{\bibfnamefont{P.}~\bibnamefont{Voisin}},
  \bibinfo{journal}{Phys. Rev. Lett.} \textbf{\bibinfo{volume}{94}},
  \bibinfo{pages}{147401} (\bibinfo{year}{2005}).    
  
\bibitem[{\citenamefont{Mei et~al.}(2005)\citenamefont{Mei, Du, Wang,
  Ying, Zeng, Zheng, Jia, Xue, and Zhang}}]{Du}
\bibinfo{author}{\bibfnamefont{Z.~X.}~\bibnamefont{Mei}},
  \bibinfo{author}{\bibfnamefont{X.~L.}~\bibnamefont{Du}},
  \bibinfo{author}{\bibfnamefont{Y.}~\bibnamefont{Wang}},
  \bibinfo{author}{\bibfnamefont{M.~J.}~\bibnamefont{Ying}},
  \bibinfo{author}{\bibfnamefont{Z.~Q.}~\bibnamefont{Zeng}},       
  \bibinfo{author}{\bibfnamefont{H.}~\bibnamefont{Zheng}},
  \bibinfo{author}{\bibfnamefont{J.~F.}~\bibnamefont{Jia}},
  \bibinfo{author}{\bibfnamefont{Q.~K.}~\bibnamefont{Xue}}, \bibnamefont{and}
  \bibinfo{author}{\bibfnamefont{Z.}~\bibnamefont{Zhang}},
  \bibinfo{journal}{Appl. Phys. Lett.} \textbf{\bibinfo{volume}{86}},
  \bibinfo{pages}{112111} (\bibinfo{year}{2005}). 
     
\bibitem[{Pui()}]{Puissance}
\bibinfo{note}{Similar effects were observed with excitation powers in the
  range $P_{exc}$=0.1 - 5 mW}. 
     
\bibitem[{\citenamefont{Reynolds et~al.}(1998)\citenamefont{Reynolds,
  Look, Jogai, Litton, Collins, Harsch, and
  Cantwell}}]{Reynolds_98}
\bibinfo{author}{\bibfnamefont{D.~C.}~\bibnamefont{Reynolds}},
  \bibinfo{author}{\bibfnamefont{D.~C.}~\bibnamefont{Look}},
  \bibinfo{author}{\bibfnamefont{B.} \bibnamefont{Jogai}},
  \bibinfo{author}{\bibfnamefont{C.~W.}~\bibnamefont{Litton}},
  \bibinfo{author}{\bibfnamefont{T.~C.} \bibnamefont{Collins}},
  \bibinfo{author}{\bibfnamefont{W.} \bibnamefont{Harsch}}, \bibnamefont{and}
  \bibinfo{author}{\bibfnamefont{G.} \bibnamefont{Cantwell}},
  \bibinfo{journal}{Phys. Rev. B} \textbf{\bibinfo{volume}{57}},
  \bibinfo{pages}{12151} (\bibinfo{year}{1998}).  

\bibitem[{\citenamefont{Meyer et~al.}(2004)\citenamefont{Meyer, Alves,
  Hofmann, Kriegseis, Forster, Bertram, Christen, Hoffmann, Stra\ss burg, Dworzak
  et~al.}}]{Meyer}
\bibinfo{author}{\bibfnamefont{B.~K.}~\bibnamefont{Meyer}},
  \bibinfo{author}{\bibfnamefont{H.}~\bibnamefont{Alves}},
  \bibinfo{author}{\bibfnamefont{D.~M.}~\bibnamefont{Hofmann}},
  \bibinfo{author}{\bibfnamefont{W.}~\bibnamefont{Kriegseis}},
  \bibinfo{author}{\bibfnamefont{K.}~\bibnamefont{Forster}},
  \bibinfo{author}{\bibfnamefont{A.}~\bibnamefont{Bertram}},
  \bibinfo{author}{\bibfnamefont{Y.}~\bibnamefont{Christen}},
  \bibinfo{author}{\bibfnamefont{G.}~\bibnamefont{Hoffmann}},
  \bibinfo{author}{\bibfnamefont{R.}~\bibnamefont{Stra\ss burg}},
  \bibinfo{author}{\bibfnamefont{C.}~\bibnamefont{Dworzak}},
  \textit{\bibnamefont{et~al.}}, \bibinfo{journal}{Phys. Stat. Sol. b}
  \textbf{\bibinfo{volume}{241}}, \bibinfo{pages}{231}
  (\bibinfo{year}{2004}). 

\bibitem[{Mag()}]{Magneto}
\bibinfo{note}{Magneto-PL data, which will be published elsewhere, confirm this assignment}.

\bibitem[{Int()}]{Integration}
\bibinfo{note}{The time integration has been performed for t$>$60 ps to avoid
  the backscattered laser light from the sample surface}.
    
\bibitem[{\citenamefont{Ishiguro et~al.}(2007)\citenamefont{Ishiguro,
  Toda, and Adachi}}]{Ishiguro}
\bibinfo{author}{\bibfnamefont{T.}~\bibnamefont{Ishiguro}},
  \bibinfo{author}{\bibfnamefont{Y.} \bibnamefont{Toda}},
  \bibnamefont{and} \bibinfo{author}{\bibfnamefont{A.}~\bibnamefont{Adachi}},
  \bibinfo{journal}{Appl. Phys. Lett.} \textbf{\bibinfo{volume}{90}},
  \bibinfo{pages}{011904} (\bibinfo{year}{2007}).
  
\bibitem[{\citenamefont{Brimont et~al.}(2004)\citenamefont{Brimont,
  Gallart, Cr\'egut, H\"onerlage, and Gilliot}}]{Gilliot}
\bibinfo{author}{\bibfnamefont{C.}~\bibnamefont{Brimont}},
  \bibinfo{author}{\bibfnamefont{M.} \bibnamefont{Gallart}},
    \bibinfo{author}{\bibfnamefont{O.} \bibnamefont{Cr\'egut}},
    \bibinfo{author}{\bibfnamefont{B.} \bibnamefont{H\"onerlage}},
    \bibnamefont{and} \bibinfo{author}{\bibfnamefont{P.}~\bibnamefont{Gilliot}},
  \bibinfo{journal}{Phys. Rev. B} \textbf{\bibinfo{volume}{77}},
  \bibinfo{pages}{125201} (\bibinfo{year}{2008}).  
  
\bibitem[{Rel()}]{Relaxation}
\bibinfo{note}{For simplicity, we neglect single particule (electron or hole) spin
relaxation processes within the free exciton with respect to exciton pseudo-spin-flip as a whole}.
 
\bibitem[{Ion()}]{Ionization}
\bibinfo{note}{We neglect the exchange energy splitting between J=1 and J=2 exciton states yielding the same ionization time $\tau_{ion}$ to those states. We have checked that the final results are not impacted in any case}.
  
\bibitem[{\citenamefont{Baylac et~al.}(1995)\citenamefont{Baylac, Marie,
  Amand, Brousseau, Barrau, and Shekun}}]{Baylac}
\bibinfo{author}{\bibfnamefont{B.}~\bibnamefont{Baylac}},
  \bibinfo{author}{\bibfnamefont{X.}~\bibnamefont{Marie}},
  \bibinfo{author}{\bibfnamefont{T.}~\bibnamefont{Amand}},
  \bibinfo{author}{\bibfnamefont{M.}~\bibnamefont{Brousseau}},
  \bibinfo{author}{\bibfnamefont{J.}~\bibnamefont{Barrau}}, \bibnamefont{and}
  \bibinfo{author}{\bibfnamefont{Y.}~\bibnamefont{Shekun}},
  \bibinfo{journal}{Surface Science} \textbf{\bibinfo{volume}{326}},
  \bibinfo{pages}{161} (\bibinfo{year}{1995}).  
    
 
\bibitem[{\citenamefont{Morhain et~al.}(2002)\citenamefont{Morhain, Teisseire,
  V\'ezian, Vigu\'e, Raymond, Lorenzini, Guion, Neu, and Faurie}}]{Morhain}
\bibinfo{author}{\bibfnamefont{C.}~\bibnamefont{Morhain}},
  \bibinfo{author}{\bibfnamefont{M.}~\bibnamefont{Teisseire}},
  \bibinfo{author}{\bibfnamefont{S.}~\bibnamefont{V\'ezian}},
  \bibinfo{author}{\bibfnamefont{F.}~\bibnamefont{Vigu\'e}},
  \bibinfo{author}{\bibfnamefont{F.}~\bibnamefont{Raymond}},
  \bibinfo{author}{\bibfnamefont{P.}~\bibnamefont{Lorenzini}},
  \bibinfo{author}{\bibfnamefont{J.}~\bibnamefont{Guion}},
  \bibinfo{author}{\bibfnamefont{G.}~\bibnamefont{Neu}}, \bibnamefont{and}
  \bibinfo{author}{\bibfnamefont{J.-P.}~\bibnamefont{Faurie}},
  \bibinfo{journal}{Phys. Stat. Sol. b} \textbf{\bibinfo{volume}{229}},
  \bibinfo{pages}{881} (\bibinfo{year}{2002}).  
   
\bibitem[{\citenamefont{Gutowski et~al.}(1988)\citenamefont{Gutowski, Presser, 
  and Broser}}]{Gutowski}
\bibinfo{author}{\bibfnamefont{J.}~\bibnamefont{Gutowski}},
  \bibinfo{author}{\bibfnamefont{N.}~\bibnamefont{Presser}}, \bibnamefont{and}
  \bibinfo{author}{\bibfnamefont{I.}~\bibnamefont{Broser}},
  \bibinfo{journal}{Phys. Rev. B} \textbf{\bibinfo{volume}{38}},
  \bibinfo{pages}{9746} (\bibinfo{year}{1988}).  
  
\bibitem[{\citenamefont{Sham et~al.}(1988)\citenamefont{Uenoyama and Sham}}]{Sham}
\bibinfo{author}{\bibfnamefont{T.}~\bibnamefont{Uenoyama}},  \bibnamefont{and}
  \bibinfo{author}{\bibfnamefont{L.~J.}~\bibnamefont{Sham}},
  \bibinfo{journal}{Phys. Rev. B} \textbf{\bibinfo{volume}{42}},
  \bibinfo{pages}{7114} (\bibinfo{year}{1990}).  
  
\bibitem[{\citenamefont{Barrau et~al.}(1993)\citenamefont{Barrau, Bacquet, Hassen, Lauret, Amand, 
  and Brousseau}}]{Barrau}
\bibinfo{author}{\bibfnamefont{J.}~\bibnamefont{Barrau}},
\bibinfo{author}{\bibfnamefont{G.}~\bibnamefont{Bacquet}},
\bibinfo{author}{\bibfnamefont{F.}~\bibnamefont{Hassen}},
\bibinfo{author}{\bibfnamefont{N.}~\bibnamefont{Lauret}},
\bibinfo{author}{\bibfnamefont{T.}~\bibnamefont{Amand}}, \bibnamefont{and}
  \bibinfo{author}{\bibfnamefont{M.}~\bibnamefont{Brousseau}},
  \bibinfo{journal}{Superlattices microstruct.} \textbf{\bibinfo{volume}{14}},
  \bibinfo{pages}{27} (\bibinfo{year}{1993}).    


\end{thebibliography}
\end{document}